    \documentclass[12pt,psf,epsf]{article}
\usepackage[centertags]{amsmath}
\usepackage{latexsym}
\usepackage{mathtools}
\usepackage{amssymb}
\usepackage{graphicx}
\usepackage{epsfig}
\usepackage{ulem}
\usepackage[english]{babel}
\usepackage{array}
\usepackage{amsthm}
\usepackage{latexsym}
\usepackage[mathcal]{euscript}
\pdfoutput=1
\usepackage{epsfig}

 \usepackage{jheppub}
 \usepackage{hyperref}

\usepackage{subfig}
\usepackage{psfrag}

\usepackage[latin1]{inputenc}
\usepackage{float}
\usepackage{cancel}
\usepackage{mathrsfs}
\usepackage{amssymb}
\usepackage{amsfonts}
\usepackage{amsmath}
\usepackage{slashed}
\usepackage{bm}
\usepackage{color}
\usepackage{appendix}
\newcommand{\be}{\begin{equation}}
\newcommand{\ee}{\end{equation}}
\newcommand{\bea}{\begin{eqnarray}}
\newcommand{\eea}{\end{eqnarray}}
\newcommand{\bwt}{\begin{widetext}}
\newcommand{\ewt}{\end{widetext}}

\newcommand{\bi}{\begin{itemize}}
\newcommand{\ei}{\end{itemize}}

\usepackage{setspace}

\definecolor{dgreen}{rgb}{0.,0.6,0.}

\begin{document}

\title{From Confinement to Chaos in AdS/CFT Correspondence via Non-equilibrium Local States}

\author{Dmitry S. Ageev$^{a,b}$, Vladimir A. Bykov$^{a,b}$}
\affiliation{$^a$Steklov Mathematical Institute, Russian Academy of Sciences, Gubkin str. 8, 119991
Moscow, Russia\\
$^b$Institute for Theoretical and Mathematical Physics, Lomonosov Moscow State University, 119991 Moscow, Russia}
\emailAdd{ageev@mi-ras.ru, bykov.va20@physics.msu.ru}

\abstract{In this paper, we study excited states in Anti-de Sitter (AdS) space prepared by local operator insertions of a massive scalar field, corresponding to local operator quenches in a free bulk scalar theory. Using the AdS/CFT correspondence, we compute the time evolution of boundary observables in the dual CFT states. We then introduce a hard wall in AdS Poincare coordinates to impose an infrared cutoff (hard-wall), creating a confining deformation of the dual conformal field theory, and analyze the dynamics of excited states in this confining background. By comparing the evolution of boundary two-point correlation functions in the deformed theory to the statistics of Gaussian random matrix ensembles, we show that for sufficiently heavy operators, the spacing-ratio statistics of peaks in temporal dynamics are closest to those of the Gaussian Symplectic Ensemble (GSE). Finally, we extend the analysis to the compact BTZ black hole and to its hard-wall deformation, finding qualitatively similar trends.}

\maketitle

\newpage

\section{Introduction} \label{sec:intro}

The study of complex quantum systems through multidisciplinary approaches has profoundly advanced our understanding of fundamental physical phenomena. Holography stands out as a powerful paradigm in this endeavor~\cite{Maldacena:1997re, Witten:1998qj}. It bridges quantum field theory, gravity, and condensed matter physics~\cite{Hartnoll:2018xxg, Zaanen:2015oix} by mapping strongly coupled conformal field theories (CFTs) to weakly coupled gravitational theories in Anti-de Sitter (AdS) spacetime. 

This duality has shed light on intricate dynamics in out-of-equilibrium states. Quantum chaos manifests there through rapid scrambling of information and thermalization, reminiscent of black hole physics~\cite{Sekino:2008he, Shenker:2013pqa, Ageev:2020qox, Rosenhaus:2020tmv, Ageev:2021poy, Gross:2021gsj}. Recent advancements underscore universal signatures, including exponential growth in out-of-time-order correlators~\cite{Maldacena:2015waa} and spectral statistics akin to random matrix theory~\cite{Brandino:2010sv, Cotler:2016fpe, Srdinsek:2020bpq, Bianchi:2022mhs, Bianchi:2023uby}.

These traits appear in holographic models and many-body quantum systems, such as Sachdev-Ye-Kitaev models or disordered spin chains~\cite{ Maldacena:2016hyu, Liu:2020rrn, Bunkov:2010fq} or in superfluid current of spins~\cite{  Bunkov:2010fq}. Also, notice holographic studies of quantum scars \cite{Milekhin:2023was}. Essential to these explorations are local operator quenches~\cite{Calabrese:2006rx, Calabrese:2007rg, Calabrese:2007mtj, Das:2014jna, Das:2014hqa, Calabrese:2016xau, Sotiriadis:2009fdq, Nozaki:2014hna, Ageev:2023wrb, Ageev:2023gna, Ageev:2017wet, Ageev:2018nye, Ageev:2018tpd, Ageev:2022kpm}. They disrupt the system via insertion of a localized operator, initiating non-equilibrium evolution. This process examines relaxation timescales, entanglement propagation, and distinctions between chaotic and integrable regimes in finite-volume or gapped theories.  In \cite{Ageev:2025yiq} we found that local quenches in massive free field theories exhibit chaotic signatures in finite volume.  In this paper we bring these results to a holographic setting, namely to confining AdS backgrounds.  

In high-energy physics, confining theories---like quantum chromodynamics (QCD) at low energies---display color confinement. Here, quarks and gluons form hadrons under a linear potential at large separations. This generates a mass gap and discrete spectrum, which curbs long-range correlations and modifies thermalization relative to gapless counterparts. Within holography, confinement corresponds to CFTs altered by relevant operators or infrared cutoffs~\cite{Casalderrey-Solana:2011dxg, Arefeva:2014kyw, DeWolfe:2013cua}. These modifications produce dual gravitational setups with adjusted AdS boundaries or brane configurations that impose a characteristic scale. Such models replicate the confinement-deconfinement phase transition and facilitate investigations of chaotic dynamics in theories lacking scale invariance. The holographic model adopted here entails capping off AdS spacetime via a hard-wall cutoff~\cite{Polchinski:2000uf, Erlich:2005qh}. It enforces Dirichlet boundary conditions on bulk fields to emulate confinement effects. Consequently, this discretizes Kaluza-Klein modes and induces a gapped spectrum on the boundary CFT.

In the present study, we undertake a comprehensive calculation of out-of-equilibrium dynamics induced by local scalar field quenches within both pure AdS and black hole backgrounds augmented with these confining cutoffs, employing exact analytical expressions for correlation functions in the bulk and on the boundary to track the system's evolution from initial perturbation to long-time behavior. Specifically, we derive two-point correlators post-quench using Green's function methods in deformed geometries, then focus on one-point functions of quadratic composite operators---such as $\phi^2$ in the bulk or $\mathcal{O}^2$ on the boundary---to quantify the relaxation and oscillatory patterns, revealing how the quench propagates as wavefronts that reflect off the hard wall and interfere constructively or destructively over time. To probe chaotic signatures, we analyze peak spacing statistics in these temporal profiles extending recent proposal of ~\cite{Bianchi:2022mhs, Bianchi:2023uby} from string amplitudes to out-of equilibrium setting (also see~\cite{Savic:2024ock, Ghodrati:2023uef, Das:2023xge, Bhattacharya:2024szw, Pesando:2025ztr} studies in this direction), comparing the distribution of spacing ratios to predictions from random matrix theory ensembles like the Gaussian Orthogonal Ensemble (GOE), Gaussian Unitary Ensemble (GUE) or Gaussian Symplectic Ensemble (GSE), thereby establishing quantitative measures of level repulsion and ergodicity. Our findings demonstrate a marked amplification of chaos with increasing scalar masses (corresponding to larger conformal dimensions $\Delta$) or when the confining wall is positioned closer to the boundary, which intensifies the infrared truncation and enhances nonlinear interactions among modes, leading to stronger deviations from Poissonian statistics toward Wigner-Dyson distributions indicative of quantum chaos. 


These results not only elucidate the interplay between confinement, mass gaps, and chaos in holographic settings but also provide benchmarks for comparing with lattice simulations of confining gauge theories or experimental realizations in quantum simulators, potentially guiding future explorations of thermalization in isolated quantum systems with tunable gaps.

The paper is structured as follows: Section 2 elaborates on local quenches in pure AdS, deriving correlation functions and boundary observables; Section 3 generalizes to capped-off AdS, scrutinizing chaotic traits through spectral statistics; Section 4 investigates quenches in BTZ black holes, encompassing both undeformed and deformed cases, alongside numerical assessments of dynamics; appendices furnish detailed derivations of Green's functions for the modified geometries.

\section{Local operator quenches in AdS}
\label{sec:pure_ads}
We consider massive scalar field theory with the action
\be
S=\frac{1}{2}\int{ d^{d+1}x\sqrt{|g|}\left[g^{\mu\nu}\partial_{\mu}\phi \partial_{\nu}\phi+m^2\phi^2\right]} \label{action}
\ee
in $AdS_{d+1}$ geometry  and for  simplicity let us start with it in  the Poincare coordinates. The metric of $AdS_{d+1}$ in Poincare coordinates (Euclidean version)  dual to $d-$dimensional  CFT is given by
\be 
ds^2=\frac{L^2}{z^2}\left(d\tau^2 +dx^2 +dz^2 \right),
\ee 
where $x=(x^1,...,x^{d-1})$, while the Lorentzian metric is obtained by Wick rotation $\tau\rightarrow it$.
Two-point correlation function of massive scalar field in such geometry has the form \cite{Allen:1985wd}
\be 
\langle\phi(\tau_1,x_1,z_1) \phi(\tau_2,x_2,z_2) \rangle = \frac{\Gamma(\Delta)}{2L^{d-1}\pi^{\frac{d}{2}}\Gamma(\Delta-\frac{d}{2}+1)}\left(\frac{\sigma_{12}}{2}\right)^{\Delta}{_2}F_1(a,b;c;\sigma_{12}^2),
\ee 
where $\Delta$, parameters $a$, $b$, $c$ and geodesic distance $\sigma_{12}$ are given by
\begin{gather}
\Delta=\frac{d}{2}+\nu, \label{Delta} \quad
\nu=\sqrt{m^2L^2+\frac{d^2}{4}},  \\
a=\frac{\Delta}{2}, \quad b=\frac{\Delta+1}{2}, \quad c=\Delta-\frac{d}{2}+1,\\
\sigma_{12}=\frac{2z_1z_2}{z_1^2+z_2^2+(x_1-x_2)^2+(\tau_1-\tau_2)^2}. \label{sigma12 poincare ads} 
\end{gather}
Following \cite{Nozaki:2014hna} we define local quench states $|\Psi(\tau)\rangle$ given by the insertion of operator $O$ at spacetime point $(\tau_q,x_q,z_q)$
\be
|\Psi(\tau)\rangle=\mathcal{N}_0\cdot e^{-H(\tau-\tau_q)}\cdot e^{-\epsilon H}O(\tau_q,x_q,z_q)|0\rangle,
\ee
so that the expectation value of a local operator 
$A$ at Euclidean time $\tau$ is given by
\be 
\langle A(\tau,x,z)\rangle_{O}=\frac{\langle0|O(-\epsilon+\tau_q,x_q,z_q)A(\tau,x,z)O(\epsilon+\tau_q,x_q,z_q)|0\rangle}{\langle0|O(-\epsilon+\tau_q,x_q,z_q)O(\epsilon+\tau_q,x_q,z_q)|0\rangle},
\ee 
where  $\epsilon$ is the regulator controlling the damping of divergent UV modes. Here and below we continue to real time via $\tau \rightarrow i  t$ (for the detailed explanation see appendix A of~\cite{Ageev:2022kpm}). The boundary dynamics corresponding to  such a state could be defined via the BDHM prescription \cite{Banks:1998dd} (see \cite{Keranen:2014lna} for AdS/CFT dictionary out of equilibrium) pushing the correlation function in the bulk to the boundary via formula
\be 
 \langle \mathcal{O}(\tau_1,x_1)\mathcal{O}(\tau_2,x_2)\rangle=\lim_{z \to 0}z^{-2\Delta}\langle\phi(\tau_1,x_1,z)\phi(\tau_2,x_2,z)\rangle_\text{O}.
\ee 
The natural candidate for the quench operator is just operator $\phi$ known to be dual to primary field $\mathcal{O}$ in $d$-dimensional CFT defined on the boundary. Then bulk two-point correlator of $\phi$   after local quench is given by
\be
\langle\phi(\tau_1,x_1,z_1)\phi(\tau_2,x_2,z_2)\rangle_{\phi}=\frac{\langle 0|\phi(-\epsilon+\tau_q,x_q,z_q)\phi(\tau_1,x_1,z_1)\phi(\tau_2,x_2,z_2)\phi(\epsilon+\tau_q,x_q,z_q)|0\rangle}{\langle 0|\phi(-\epsilon+\tau_q,x_q,z_q)\phi(\epsilon+\tau_q,x_q,z_q)|0\rangle}.
\ee
Now the comment is in order.    We are going to perform calculations in free scalar field theory in AdS in the context of excited state which effectively reduces to four-point correlation function in the bulk. At leading order in large $N$, correlators of the single-trace operators dual to our free bulk scalar are generalized free; the four-point function that appears in the normalized quenched two-point is therefore dominated by its disconnected part with the subleading connected pieces being of order $\mathcal{O}\left(1 / N^2\right)$ and  neglected.

Thus we consider free theory, and  express correlation function after the quench in the bulk explicitly as the sum of three Wick contractions 
\begin{multline}
\label{two point quench poincare ads}
\langle\phi(\tau_1,x_1,z_1)\phi(\tau_2,x_2,z_2)\rangle_{\phi}=\frac{\Gamma(\Delta)}{2\pi^{\frac{d}{2}} L^{d-1}\Gamma(\Delta-\frac{d}{2}+1)}\left(\frac{\sigma_{12}}{2}\right)^{\Delta}{_2}F_1(a,b;c;\sigma_{12}^2) +\\
    +\frac{\Gamma(\Delta)}{2\pi^{\frac{d}{2}} L^{d-1}\Gamma(\Delta-\frac{d}{2}+1)}\left(\frac{\sigma_1^+\sigma_2^-}{2\sigma_q}\right)^{\Delta}\frac{_2F_1(a,b;c;(\sigma_1^+)^2)\cdot {_2}F_1(a,b;c;(\sigma_2^-)^2)}{_2F_1(a,b;c;\sigma_q^2)}+\\
    +\frac{\Gamma(\Delta)}{2\pi^{\frac{d}{2}} L^{d-1}\Gamma(\Delta-\frac{d}{2}+1)}\left(\frac{\sigma_1^-\sigma_2^+}{2\sigma_q}\right)^{\Delta}\frac{_2F_1(a,b;c;(\sigma_1^-)^2)\cdot{_2}F_1(a,b;c;(\sigma_2^+)^2)}{_2F_1(a,b;c;\sigma_q^2)},
\end{multline}
first of which is just the two-point correlation function without any quench, and the others control the dynamics in the bulk after the quench. Here $\sigma_{12}$ is given by \eqref{sigma12 poincare ads}, while $\sigma_q$,  $\sigma_1^{\pm}$ and $\sigma_2^{\pm}$   are defined as 
\begin{equation}
    \sigma_q=\frac{z_q^2}{z_q^2+2\epsilon^2},\,\,\,\,
\sigma_{1,2}^{\pm}=\frac{2z_{1,2}z_q}{z_{1,2}^2+z_q^2+(\tau_{1,2}\pm \epsilon-\tau_q)^2+(x_{1,2}-x_q)^2}. \label{sigma pm 12 poincare ads}
\end{equation}
After applying BDHM rule we are left with the boundary correlation function corresponding to the quantum out-of-equilibrium bulk quench state 
\begin{multline}
\label{two point quench CFT poincare AdS}
    \langle \mathcal{O}(\tau_1,x_1)\mathcal{O}(\tau_2,x_2)\rangle_{\phi}=\frac{\Gamma(\Delta)}{2\pi^{\frac{d}{2}} L^{d-1}\Gamma(\Delta-\frac{d}{2}+1)[(\tau_1-\tau_2)^2+(x_1-x_2)^2]^{\Delta}}+\\+\frac{\Gamma(\Delta)((\eta_1^+\eta_2^-)^{\Delta}+(\eta_1^-\eta_2^+)^{\Delta})}{2\pi^{\frac{d}{2}} L^{d-1}\Gamma(\Delta-\frac{d}{2}+1)(2\sigma_q)^{\Delta}\cdot{_2}F_1(a,b;c;\sigma_q^2)},
\end{multline}
where one explicitly sees that the first term is just the equilibrium two-point correlation function of primary operators as it should be and $\eta_{1,2}^{\pm}$ is given by
\begin{equation}
    \eta_{1,2}^{\pm}=\frac{2z_q}{z_q^2+(\tau_{1,2}\pm \epsilon-\tau_q)^2+(x_{1,2}-x_q)^2}.
\end{equation}
For simplicity, in this paper we focus on the dynamics of a one-point correlator of the composite operator $\phi^2(\tau,x,z)$ and $\langle\mathcal{O}^2(\tau,x)\rangle_{\phi}$ to avoid issues related to analytic continuation. Often in such problems, the dynamics of an equal-time two-point correlation function is studied, and we checked that it resembles $\phi^2$ dynamics. After taking the coincident points limit of (\ref{two point quench poincare ads}) and subtracting the divergencies in a minimal way we end up with 
\be
\langle\phi^2(t,x,z)\rangle_{\phi}=
    \frac{\Gamma(\Delta)}{\pi^{\frac{d}{2}} L^{d-1}\Gamma(\Delta-\frac{d}{2}+1)}\left(\frac{\sigma^+\sigma^-}{2\sigma_q}\right)^{\Delta}\frac{_2F_1(a,b;c;(\sigma^+)^2)\cdot{_2}F_1(a,b;c;(\sigma^-)^2)}{_2F_1(a,b;c;\sigma_q^2)},
\ee
where we analytically continue $\tau \rightarrow i t$ and
\be
\sigma^{\pm}=\frac{2zz_q}{z^2+z_q^2+(i t \pm \epsilon-i t_q)^2+(x-x_q)^2}.
\ee
Doing the same for (\ref{two point quench CFT poincare AdS}) we get
\be
\langle\mathcal{O}^2(t,x)\rangle_{\phi}=\frac{(\eta^+\eta^-)^{\Delta}}{\pi^{\frac{d}{2}} L(2\sigma_q)^{\Delta}\cdot{_2}F_1(a,b;c;\sigma_q^2)}.
\ee
What we obtain is that in the bulk we see a localized perturbation along the geodesics (i.e. circles) that propagates into the bulk.
\begin{figure}[t!]
\centering
\subfloat[Bulk dynamics]{\includegraphics[width=0.5\textwidth]{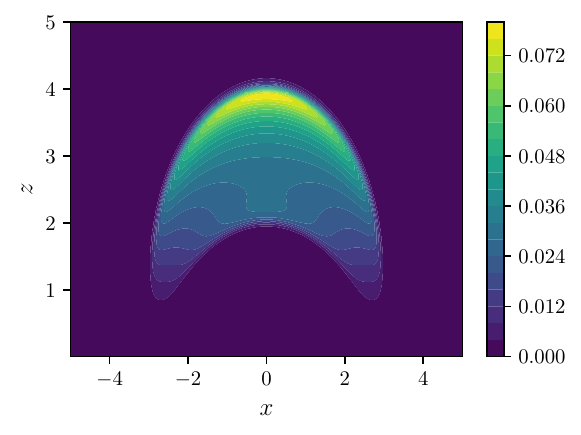}}
\subfloat[Boundary dynamics]{\includegraphics[width=0.5\textwidth]{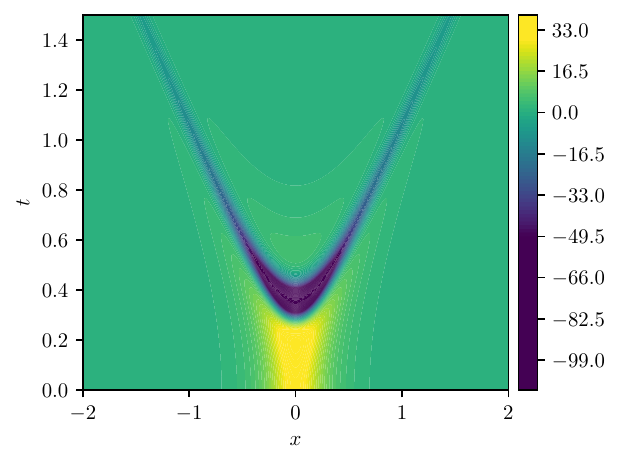}}
\caption{(a) Bulk dynamics for the condensate $\phi$ for mass parameter corresponding to $\Delta=5$, $m=3.87$ at the moment of time $t=3$ with quench point in the bulk $z_q=1$. (b) Boundary dynamics of $\mathcal{O}^2$ for mass parameter $\Delta=2$, $m=0$ with quench point in the bulk $z_q=0.35$. Parameters $d=2$, $x_q=t_q=0$, $\epsilon=0.1$, $L=1$ are fixed for all figures.} 
\label{fig:Poincare AdS_3 no wall}
\end{figure}

On the boundary it corresponds to the forming a compact pulse that rapidly splits into two localized configurations of energy counter-propagating from the quench point, i.e. in some sense it resembles the canonical local quench in the flat space. However, there are different features making difference with flat CFT quench, namely
\begin{itemize}
    \item Quench state in AdS depends less sensitively on the mass then in flat space. For all masses in quench state in AdS we observe first a fast localized increase of correlation near the quench point and then rapid decay into two compact configurations moving away from $x_q$, while in flat space for zero mass initial configuration splits and propagates freely (mass increase introduce diffusion and decrease). Also increasing the mass in AdS quench only adds some oscillations near the peaks of configurations moving away from the quench point.
    \item The parameters $z_q$ and $\epsilon$ control how deep in the infrared the quench is made and how smeared it is. Mostly, this is controlled by $z_q$. It takes some time for perturbation to get from the quench point to the bulk and in this way $z_q$ defines how fast the initial compact distribution of correlations will stop to grow in will decay into two.
\end{itemize}

\section{Capping-off AdS and chaos in confining theory}
Now deform the initial geometry of Poincare AdS by imposing a Dirichlet boundary condition on scalar field at some plane $z=z_0$ in the bulk, introducing a hard-wall IR cutoff. This corresponds to the deformation of the IR degrees of freedom in such a way that their dynamics is strongly restricted -- in other words this theory now has kind of confinement mechanism \cite{Polchinski:2000uf}. 
First, let us calculate a two-point correlation function for such theory.
For simplicity, let us introduce the following notation for boundary coordinates $y=(\tau, x)$. 

Then for massive scalar free field theory two-point correlator  $\langle \phi(y,z) \phi (y^{\prime},z^{\prime}) \rangle= G(y,z; y^{\prime},z^{\prime})$ equation of motion remains the same but boundary conditions are modified as 
\begin{gather}
    G(y,z_0;y',z')=G(y,z;y',z_0)=0 \label{boundary cond for green ads poincare},\\
    G(y,0;y',z')=G(y,z;y',0)=0. \label{boundary cond for green ads poincare at infty}
\end{gather}
The details of the calculation can be found in  Appendix \ref{App A}.

\begin{figure}[t!]
\centering
\subfloat[$t = 4$]{\includegraphics[width=0.33\textwidth]{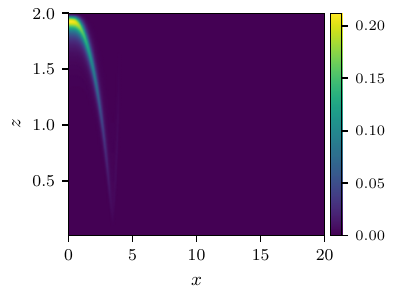}}
\subfloat[$t = 10$]{\includegraphics[width=0.33\textwidth]{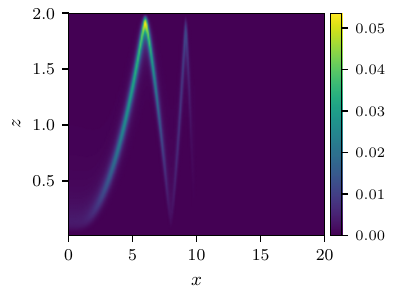}}
\subfloat[$t = 20$]{\includegraphics[width=0.33\textwidth]{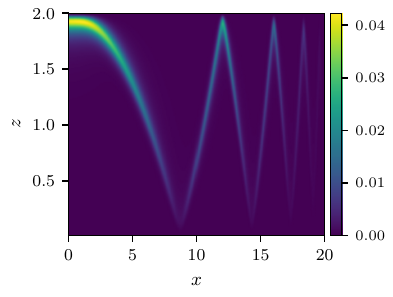}}\\
\subfloat[$t = 4$]{\includegraphics[width=0.33\textwidth]{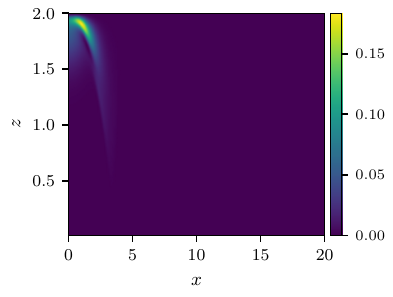}}
\subfloat[$t = 10$]{\includegraphics[width=0.33\textwidth]{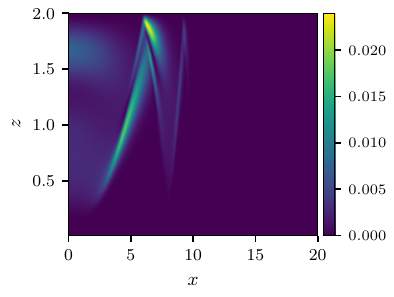}}
\subfloat[$t = 20$]{\includegraphics[width=0.33\textwidth]{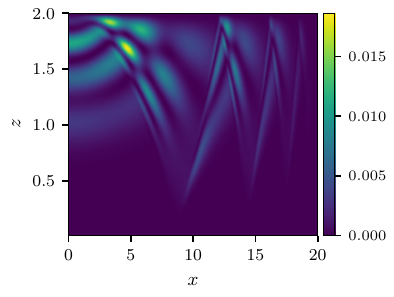}}
\caption{Bulk dynamics for the condensate $\phi$ for mass parameters corresponding $\nu=1$, $m=0$ (Top) and $\nu=7$, $m=6.93$ (Bottom). Parameters $d=2$, $z_0=2$, $x_q=t_q=0$, $z_q=1.99$, $\epsilon=0.1$, $L=1$, $N=150$ are fixed for all figures, where $N$ is the number of terms taken in the sum over n.} 
\label{fig:Poincare AdS_3}
\end{figure}

The solution for $G(y,z;y',z')$ has the form
\be
\label{ansatz for green ads poincare}
G(y,z;y',z')=\sum_n\int\frac{d^dk}{(2\pi)^d} e^{ik(y-y')}\frac{z^{\frac{d}{2}}z'^{\frac{d}{2}}J_{\nu}(\alpha_n z)J_{\nu}(\alpha_n z')}{N_n}G_n(k),
\ee
where $\alpha_n$, $N_n$, $G_n(k)$ are defined as follows
\begin{gather}
    J_{\nu}(\alpha_n z_0)=0, \label{ alpha def wall ads poincare} \\
    N_n=\int_0^{z_0}dz zJ_{\nu}^2(\alpha_n z)=\frac{z_0^2}{2}J_{\nu+1}^2(\alpha_n z_0) \label{N_n def ads poincare},\\
    G_n(k)=\frac{1}{L^{d-1}(k^2+\alpha_n^2)}
\end{gather}
and it is straightforward to see that  (\ref{ansatz for green ads poincare}) satisfies the boundary conditions (\ref{boundary cond for green ads poincare}). Performing the integration over momentum $k$ we obtain two-point correlation function defined as a sum over integers
\be
\label{green deformed ads poincare}
G(\tau,x,z;\tau',x',z')=\frac{1}{(2\pi)^{\frac{d}{2}}L^{d-1}}\sum_n\frac{z^{\frac{d}{2}}z'^{\frac{d}{2}}J_{\nu}(\alpha_n z)J_{\nu}(\alpha_n z')}{N_n}\left(\frac{\alpha_n}{r}\right)^{\frac{d}{2}-1}K_{\frac{d}{2}-1}(\alpha_n r),
\ee
where $r=\sqrt{(\tau-\tau')^2+(x-x')^2}$, (see Appendix \ref{App A} for details). Having the Green's function for the deformed geometry we proceed to define quench dynamics as we did earlier.

To obtain deformed boundary correlation function, first we apply the BDHM rule for the deformed bulk correlator (\ref{green deformed ads poincare}) to obtain
\be
G_{CFT}(\tau,x;\tau',x')=\frac{1}{(2\pi)^{\frac{d}{2}}L^{d-1}}\sum_n\frac{\alpha_n^{2\nu}}{\Gamma^2(\nu+1)2^{2\nu}N_n}\left(\frac{\alpha_n}{r}\right)^{\frac{d}{2}-1}K_{\frac{d}{2}-1}(\alpha_n r).
\ee
To simplify the notation for deformed boundary quench dynamics let us define $\tilde{G}_{CFT}$ as
\be
\tilde{G}_{CFT}(\tau,x;\tau',x',z)=\frac{1}{(2\pi)^{\frac{d}{2}}L^{d-1}}\sum_n\frac{z^{\frac{d}{2}}J_{\nu}(\alpha_n z)\alpha_n^{\nu}}{\Gamma(\nu+1)2^{\nu}N_n}\left(\frac{\alpha_n}{r}\right)^{\frac{d}{2}-1}K_{\frac{d}{2}-1}(\alpha_n r),
\ee
where $z$-dependence  corresponds to inserting the quench operator at the bulk point. Then the deformed boundary two-point correlation function after the local quench is given by
\begin{multline}
   \langle \mathcal{O}(\tau_1,x_1)\mathcal{O}(\tau_2,x_2)\rangle_{D\phi}=G_{CFT}(\tau_1,x_1;\tau_2,x_2)+\\
   \frac{\tilde{G}_{CFT}(\tau_1,x_1; -\epsilon+\tau_q,x_q,z_q)\tilde{G}_{CFT}(\tau_2,x_2;\epsilon+\tau_q,x_q,z_q)}{G(-\epsilon+\tau_q,x_q,z_q;\epsilon+\tau_q,x_q,z_q)}+\\
\frac{\tilde{G}_{CFT}(\tau_2,x_2;-\epsilon+\tau_q,x_q,z_q)\tilde{G}_{CFT}(\tau_1,x_1;\epsilon+\tau_q,x_q,z_q)}{G(-\epsilon+\tau_q,x_q,z_q;\epsilon+\tau_q,x_q,z_q)}.
\end{multline}

\begin{figure}[t!]
\centering
\subfloat[$\nu=1, m=0$]{\includegraphics[width=0.5\textwidth]{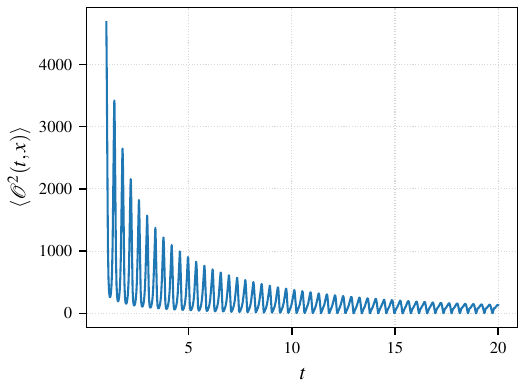}}
\subfloat[$\nu=7, m=6.93$]{\includegraphics[width=0.5\textwidth]{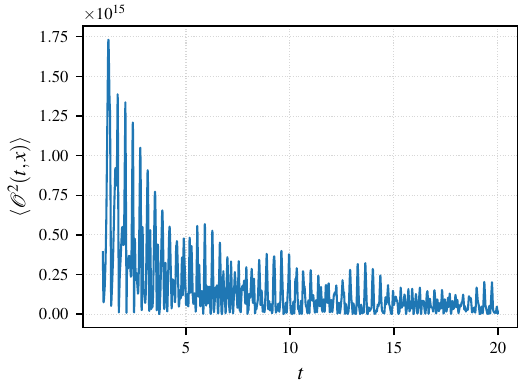}}
\caption{Boundary dynamics of primary  $\mathcal{O}^2$ for different mass parameters at point $x=0$. $d=2$, $z_0=0.2$, $x_q=t_q=0$, $z_q=1.999$, $\epsilon=0.1$, $L=1$, $N=100$ are fixed for all figures.} 
\label{fig:Boundary dynamics}
\end{figure}

The finite part of the boundary one-point correlator (Wick rotated $\tau\rightarrow it$) after the quench is written as
\be
\label{CFT one-point ads poincare}
\langle \mathcal{O}^2(t,x)\rangle_{D\phi}=
   \frac{2\tilde{G}_{CFT}(it,x; -\epsilon+it_q,x_q,z_q)\tilde{G}_{CFT}(it,x;\epsilon+it_q,x_q,z_q)}{G(-\epsilon+it_q,x_q,z_q;\epsilon+it_q,x_q,z_q)}.
\ee
To analyze the chaotic behavior of (\ref{CFT one-point ads poincare}) let us examine the statistical properties of peak spacing ratios in correlation functions following a local quench. This approach uses random matrix theory as a framework to analyze chaotic quantum dynamics. The probability density function (PDF) of spacing ratio $r_n$, defined as 
\be
r_n=\frac{\delta_{n+1}}{\delta_n},
\ee
should be compared with that of Gaussian ensembles of random matrices (GUE, GOE and GSE), where $\delta_n=\lambda_{n+1}-\lambda_n$ is spacing of peaks $\lambda_n$. Consideration of $r_n$ is known to be less sensitive to folding/unfolding procedures which is intrinsic to consideration of level spacing itself \cite{Oganesyan:2007wpd}. We provide the comparison with main Gaussian ensembles distribution of ratios, which are given by \cite{Atas:2013gvn}
\be
f_\beta(r)=\frac{3^{\frac{3+3 \beta}{2}} \Gamma\left(1+\frac{\beta}{2}\right)^2}{2 \pi \Gamma(1+\beta)} \frac{\left(r+r^2\right)^\beta}{\left(1+r+r^2\right)^{1+\frac{3}{2} \beta}}
\ee
with GOE corresponding to $\beta=1$, GUE to $\beta=2$ and GSE to $\beta=4$.

\begin{figure}[t!]
\centering
\subfloat[$\nu=1, m=0$]{\includegraphics[width=0.5\textwidth]{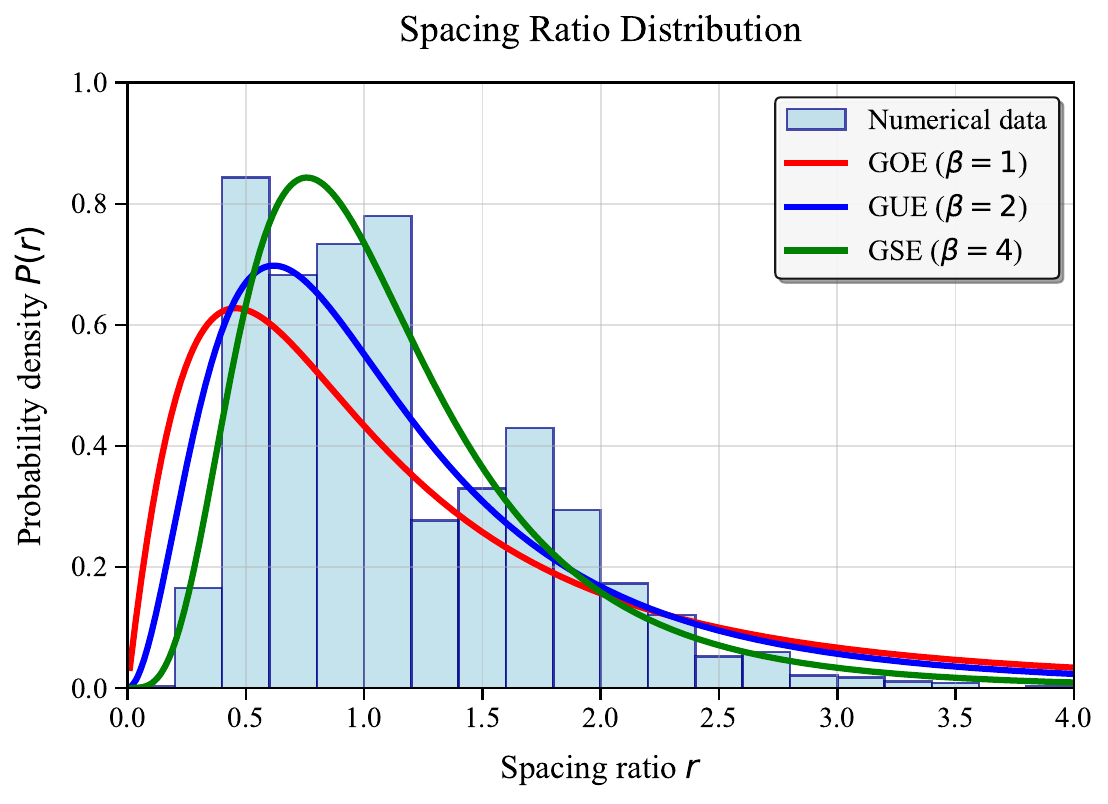}}
\subfloat[$\nu=7, m=6.93$]{\includegraphics[width=0.5\textwidth]{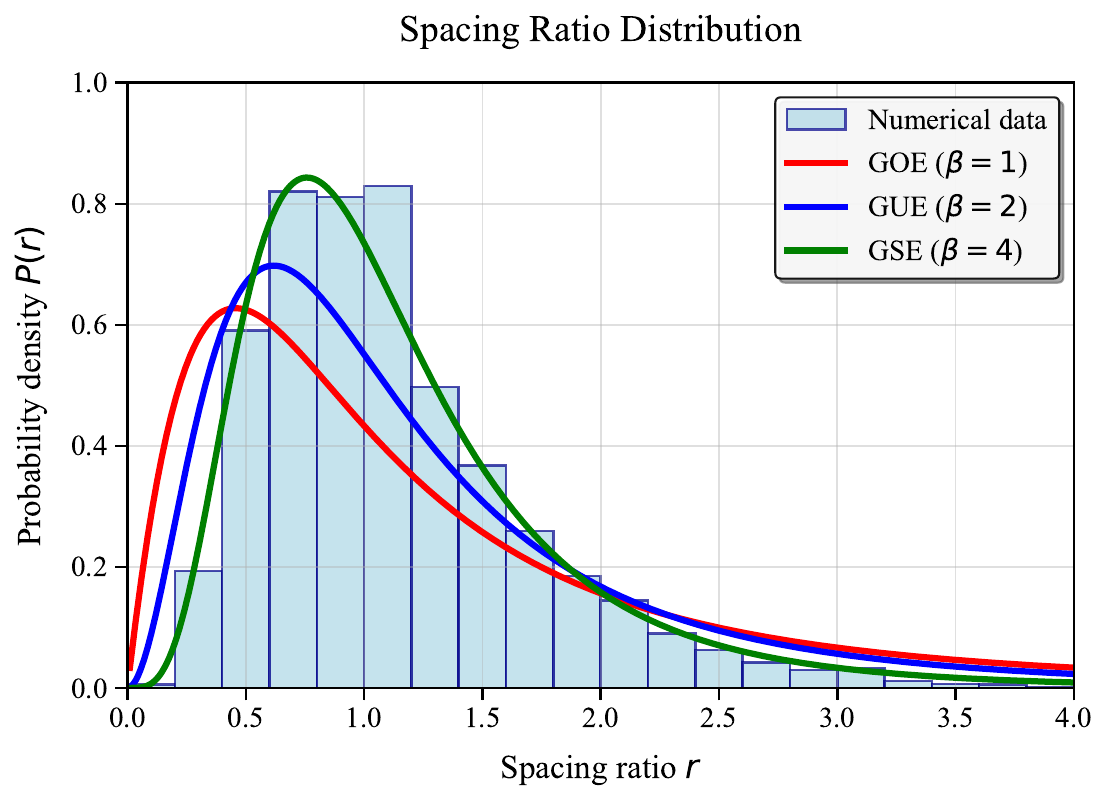}}
\caption{Distributions of peak spacing ratios for boundary dynamics of primary $\mathcal{O}^2$ for different mass (massless $\nu=1$ and large mass $\nu=7$) parameters. $d=2$, $z_0=0.2$, $z_q=0.199$, $x_q=t_q=0$, $\epsilon=0.1$, $L=1$, $N=300$ are fixed for all figures.} 
\label{fig:CFT ratio distribution}
\end{figure}

In Fig.\ref{fig:CFT ratio distribution} we show distribution for peaks ratio of $\mathcal{O}^2(\tau,x)\rangle_{D\phi}$ for different masses in $d=2$ dimensional theory. One can clearly observe the agreement with GSE ensemble (green line)   for large field masses, while for small ones we clearly observe the absence of such dynamics. Chiral cousin of symplectic ensemble is known to be strongly related to the spectrum Dirac operator in adjoint QCD. Let us highlight important features and points concerning

\begin{itemize}
\item For small values of mass (i.e. close to $\nu=1$) we see that no desired PDF close enough to any ensemble is present. For larger masses it starts to converge to desirable distribution.
\item A characteristic feature of the chaotic dynamics is the increase in chaotic behavior as the "wall" 
 approaches the conformal boundary $z_0\rightarrow0$. This corresponds to the truncation of infrared degrees of freedom and strong deformation of the boundary conformal field theory. The more confining the theory, the more pronounced the chaotic behavior. By chaotic behavior we mean that, it tends to get closer to GSE.
 \item If the quench point is closer to the wall, the chaos is present in a more pronounced form, the fit improves, and smaller masses and fewer series terms are needed to approach the GSE PDF.
 \item An interesting fact is that for large masses and if quench point is close enough to the wall then we obtain a good fit with chaotic PDFs already for $N=5$ terms of the series.
\item  Larger $\nu$ (and thus larger $\Delta$) corresponds to a more irrelevant operator in the CFT, which suppresses low-energy fluctuations and enhances nonlinear interactions/strong coupling.
\end{itemize}

\section{Towards finite temperature regimes: quenches in capped BTZ}
After we demonstrated the matching between level spacing ratios in temporal dynamics and GSE ensemble distribution in the previous section for Poincare AdS let us move to a more complicated case. Now consider massive scalar field theory with the action (\ref{action}) in BTZ black hole  background with the metric
\be
ds^2=\frac{r^2-r_h^2}{L^2}d\tau^2+\frac{L^2}{r^2-r_h^2}dr^2+r^2d\varphi^2, \quad r>r_h, \label{BTZ metric}
\ee
which is known to be dual to 2d CFT with finite temperature $T=r_h/(2\pi L^2)$. 
Here we study the version of BTZ black hole which has periodic spatial coordinate and in general it is more intricate and complicated from the physical viewpoint than its planar cousin because being dual to a theory at finite volume and temperature simultaneously. The BTZ black hole is locally equivalent to $AdS_3$ space, so the two-point correlation function of the massive scalar field has the same local form as in $AdS_3$ space. Globally, the BTZ black hole is a discrete quotient of $AdS_3$ with equivalence relation imposed on the angular variable ($\varphi \sim \varphi+2\pi k$), where $k$ is an integer. 

$\,$

A fully general study of boundary correlation functions and statistics in the spirit of the previous section is a much more technically complicated task due to the number of free parameters,  summations requiring considerable numerical capabilities even in the compact case BTZ with confining cutoff to get stable and clear results. Curiously, the planar black hole turns out to be an even more complicated object to study numerically for an IR deformation. Here we demonstrate that  behavior resembling  the previous section could be also observed in principle and describe general trends which distributions satisfy. We would like to stress that we do not pretend here on a full-scale study of chaos in thermal ``confining'' theories  and leave a more detailed study to a future research. 

$\,$

Hence two-point correlation function in BTZ is obtained from the one of $AdS_3$ using the method of images
\be
G_{BTZ}(\tau_1,\varphi_1,r_1;\tau_2,\varphi_2,r_2)=\sum_n\frac{1}{2L\pi}\left(\frac{\sigma_n^{12}}{2}\right)^{\Delta}{_2}F_1\left(\frac{\Delta}{2},\frac{\Delta+1}{2};\Delta;(\sigma^{12}_n)^2\right),
\ee
where $\sigma^{12}_n$ is the geodesic distance corresponding to the metric (\ref{BTZ metric})  and given by
\be
\sigma^{12}_n=\left(\frac{r_1r_2\cosh{\left[\frac{r_h(\varphi_1-\varphi_2+2\pi n)}{L}\right]}}{r_h^2}-\sqrt{\left(\frac{r_1^2}{r_h^2}-1\right)\left(\frac{r_2^2}{r_h^2}-1\right)}\cos{\left(\frac{r_h(\tau_1-\tau_2)}{L^2}\right)}\right)^{-1}. \label{BTZ correlator}
\ee
Following the same logic as in previous section, we obtain bulk correlator after local quench in compact BTZ black hole as
\be
\langle\phi^2(t,\varphi,r)\rangle_{\phi, BTZ}=\frac{2G_{BTZ}(-\epsilon+it_q,\varphi_q,r_q;it,\varphi,r)G_{BTZ}(\epsilon+it_q,\varphi_q,r_q;it,\varphi,r)}{G_{BTZ}(-\epsilon+it_q,\varphi_q,r_q;\epsilon+it_q,\varphi_q,r_q)}.
\ee
To define the one-point boundary observable, we set
\be
G^{CFT}_{BTZ}(\tau_1,\varphi_1;\tau_2,\varphi_2,r)=\sum_n\frac{1}{2L\pi}\left(\frac{\xi_n^{12}}{2}\right)^{\Delta},
\ee
where $\xi_n^{12}$
\be
\xi_n^{12}=\left(\frac{r\cosh{\left[\frac{r_h(\varphi_1-\varphi_2+2\pi n)}{L}\right]}}{r_h^2}-\frac{\sqrt{\frac{r^2}{r_h^2}-1}}{r_h}\cos{\left(\frac{r_h(\tau_1-\tau_2)}{L^2}\right)}\right)^{-1}
\ee
in terms of which it defines (after extracting divergences) as

\be
\langle \mathcal{O}^2(t,\varphi)\rangle_{\phi, BTZ}=
   \frac{2G^{CFT}_{BTZ}(it,\varphi;-\epsilon+it_q,\varphi_q,r_q)G^{CFT}_{BTZ}(it,\varphi;\epsilon+it_q,\varphi_q,r_q)}{G_{BTZ}(-\epsilon+it_q,\varphi_q,z_q;\epsilon+it_q,\varphi_q,z_q)}.
\ee

Now let us deform the BTZ black hole geometry by imposing a Dirichlet boundary condition on a brick-wall surface and calculate necessary two-point correlation function $G_{DBTZ}$. The technicalities are the same as in the previous section (for more details of the derivation, see Appendix \ref{App B}). 
The solution for $G_{DBTZ}(\tau,\varphi,r;\tau',\varphi',r')$ has the form
\be
\label{solution for green BTZ}
G_{DBTZ}(\tau,\varphi,r;\tau',\varphi',r')=\sum_n\sum_J\int\frac{d\lambda}{(2\pi)^2} e^{i\lambda(\tau-\tau')}e^{iJ(\varphi-\varphi')}\frac{2L^2}{r_h\sqrt{rr'}}\frac{F_{n,J}(r)F_{n,J}(r')}{(\lambda^2+\omega_{n,J}^2)N_{n,J}},
\ee
where $J$ is an integer and $F_{n,J}(r)$, $N_{n,J}$ are defined as follows
\begin{gather}
    F_{n,J}(r)=\left(1-\frac{r_h^2}{r^2}\right)^{\alpha_n}\left(\frac{r_h^2}{r^2}\right)^{\beta}{_2}F_1\left(a_{n,J},b_{n,J};c;\frac{r_h^2}{r^2}\right), \\
     N_{n,J}=\frac{L^4}{r_h^2}\int_{0}^{z_0}(1-z)^{2\alpha_{n,J}-1}z^{\nu}{_2}F_1^2(a_{n,J},b_{n,J};c;z)dz, \label{N_n def BTZ}
\end{gather}
where $z=r_h^2/r^2$.
Parameters $a_n$, $b_n$, $c$, $\alpha_n$ and $\beta$ are given by
\begin{gather}
    a_n,b_n=\frac{1}{2}\left(1+\nu+\frac{iL^2\omega_n}{r_h}\mp\frac{iLJ}{r_h}\right),  \quad c=\Delta \label{hypergeometric param},\\
    \alpha_n=\frac{iL^2\omega_n}{2r_h}, \quad \beta=\frac{1}{4}+\frac{\nu}{2}. \label{alpha and beta}
\end{gather}
where $\omega_n$ are given as solutions of equation
\be
{_2}F_1\left(a_n,b_n;c;\frac{r_h^2}{r_0^2}\right)=0. \label{boundary eq BTZ}
\ee
Note that $\omega_{n,J}$ depends on $J$. Performing the integration over $\lambda$ we get
\be
\label{green deformed BTZ}
G_{DBTZ}(\tau,\varphi,r;\tau',\varphi',r')=\sum_n\sum_J\frac{e^{-\omega_{n,J}|\tau-\tau'|}}{4\pi\omega_{n,J}}e^{iJ(\varphi-\varphi')}\frac{2L^2}{r_h\sqrt{rr'}}\frac{F_{n,J}(r)F_{n,J}(r')}{N_{n,J}}.
\ee

\begin{figure}[t!]
\centering
\subfloat[$\nu = 1$, $r_h=1$]{\includegraphics[width=0.5\textwidth]{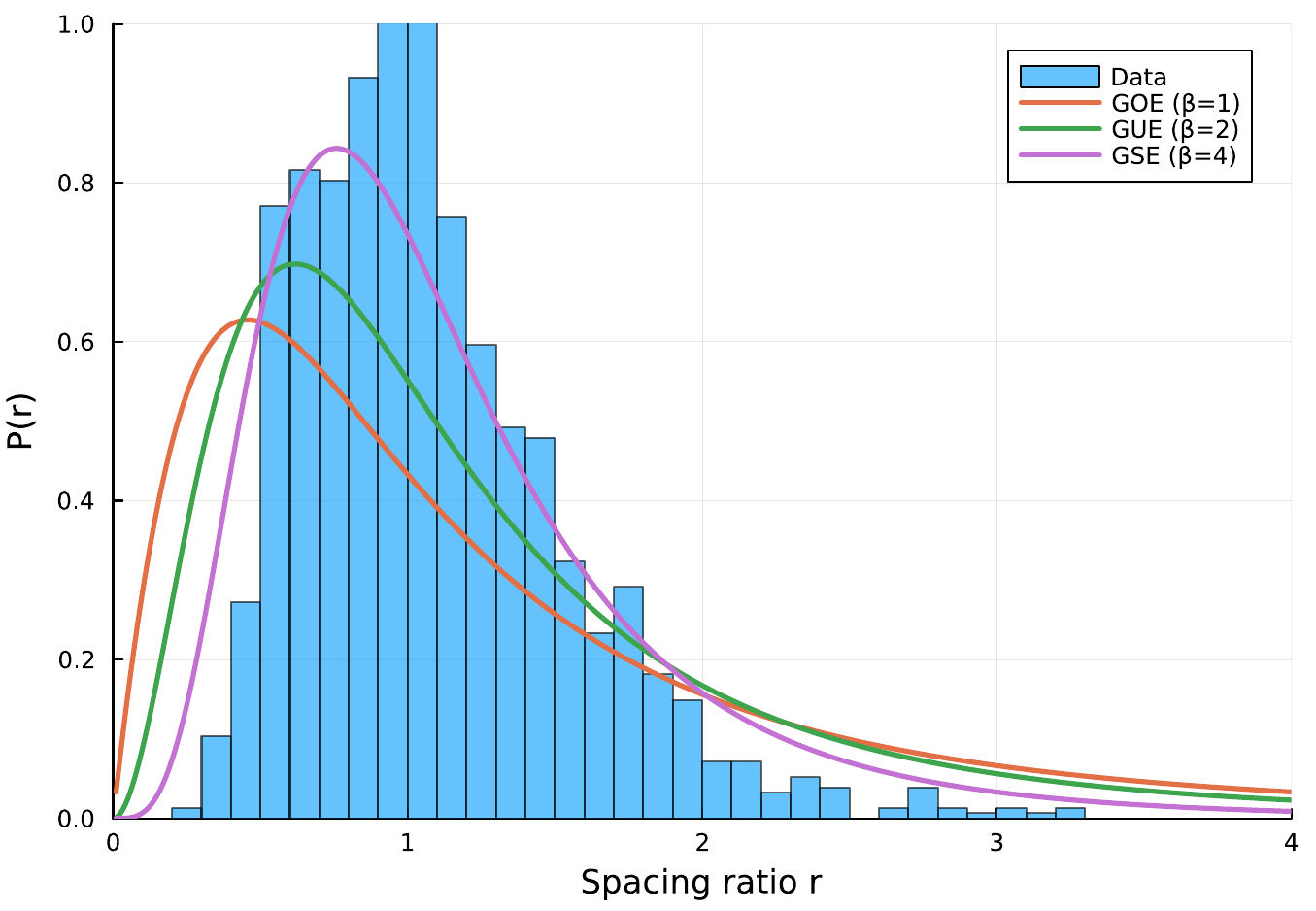}}
\subfloat[$\nu = 3$, $r_h=1$]{\includegraphics[width=0.5\textwidth]{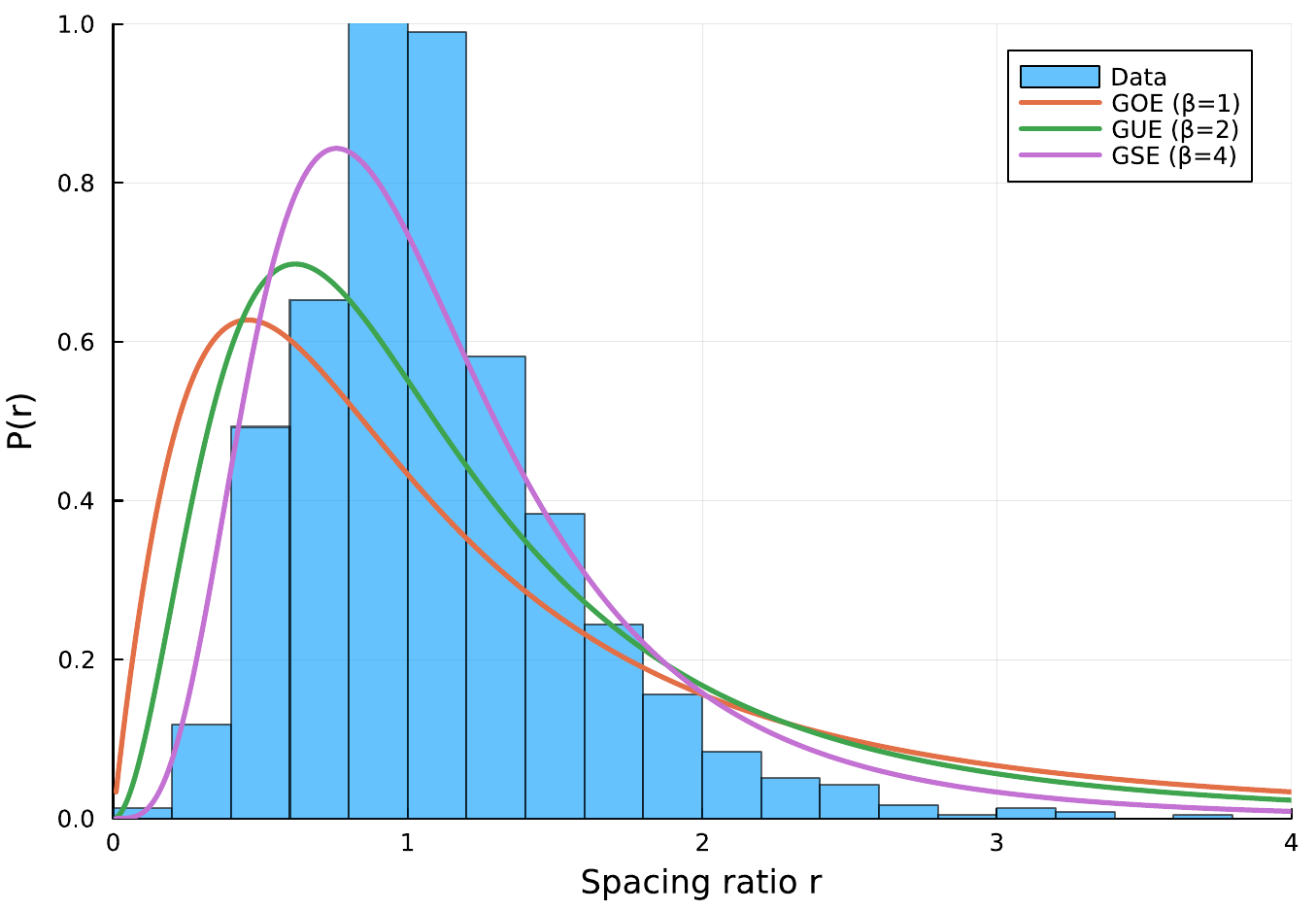}}\\
\subfloat[$\nu = 1$, $r_h=6$]{\includegraphics[width=0.5\textwidth]{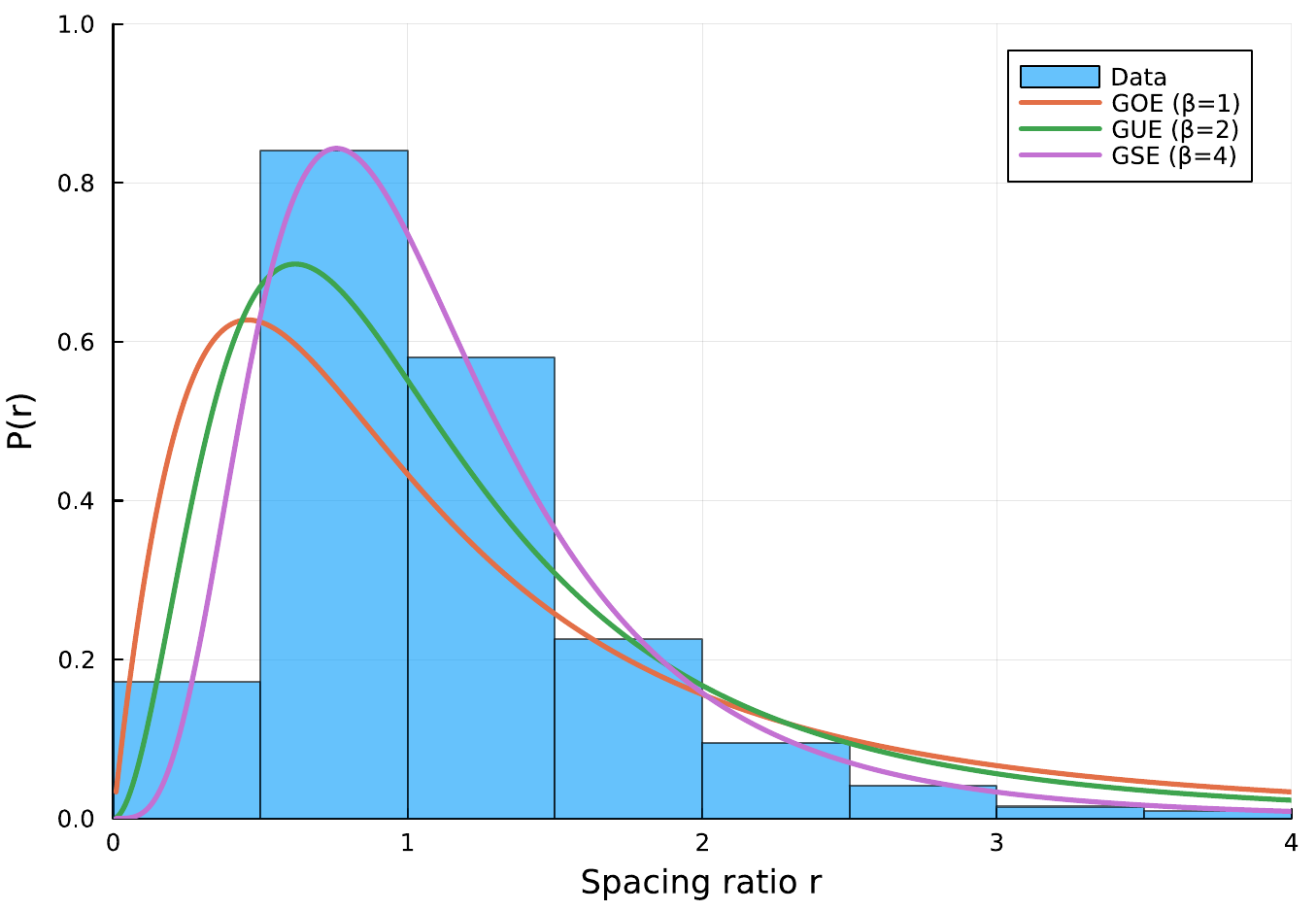}}
\subfloat[$\nu = 3$, $r_h=6$]{\includegraphics[width=0.5\textwidth]{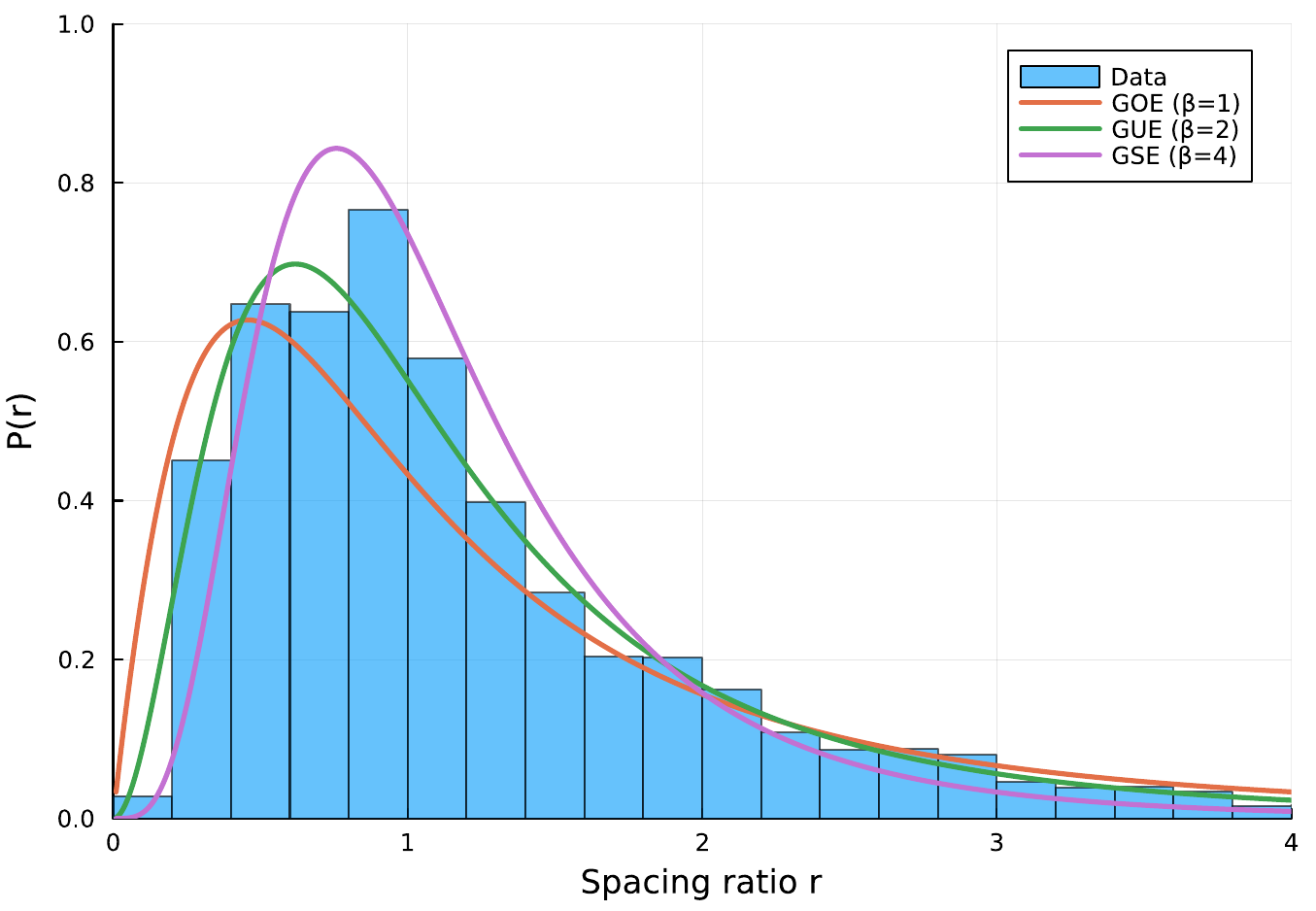}}
\caption{Distributions of peak spacing ratios for boundary dynamics of primary $\mathcal{O}^2$ for different mass (massless $\nu=1$ and large mass $\nu=3$) parameters in deformed BTZ geometry. Pannels (a)-(b) correspond to value $r_h=1$ and (c)-(d) -- $r_h=6$ (high temperature regime). Parameters $\varphi_q=t_q=0$, $\epsilon=0.1$, $L=1$, $N=25$, $J_{\text{max}}=40$, $z_0=0.99$, $z_q=0.98$  are fixed for all figures.} 
\label{fig:BTZ distribution}
\end{figure}

Having the Green function for deformed geometry we proceed to define quench dynamics as we did earlier and we define 
\be
G_{DBTZ}^{CFT}(\tau,\varphi;\tau',\varphi')=\frac{L^2r_h^{4\beta-1}}{2\pi}\sum_n\sum_J\frac{e^{-\omega_{n,J}|\tau-\tau'|}}{\omega_{n,J} N_{n,J}}e^{iJ(\varphi-\varphi')},
\ee
as well as
\be
\tilde{G}_{DBTZ}^{CFT}(\tau,\varphi;\tau',\varphi',r)=\frac{L^2r_h^{2\beta-1}}{2\pi}\sum_n\sum_J\frac{e^{-\omega_{n,J}|\tau-\tau'|}}{4\pi\omega_{n,J}}e^{iJ(\varphi-\varphi')}\frac{1}{\sqrt{r}}\frac{F_{n,J}(r)}{N_{n,J}},
\ee
then the finite part of boundary one-point quench dynamics has the form
\be
\label{CFT one-point BTZ deformed}
\langle \mathcal{O}^2(t,\varphi)\rangle_{\phi,DBTZ}=
   \frac{2\tilde{G}_{DBTZ}^{CFT}(it,\varphi;-\epsilon+it_q,\varphi_q,r_q)\tilde{G}_{DBTZ}^{CFT}(it,\varphi;\epsilon+it_q,\varphi_q,r_q)}{G_{DBTZ}(-\epsilon+it_q,\varphi_q,r_q;\epsilon+it_q,\varphi_q,r_q)}.
\ee

In Fig. \ref{fig:BTZ distribution} we show the distribution for the peaks ratio in the deformed BTZ geometry, regulated using a brick wall model with the wall placed very close to the horizon at $z_0=r_h^2/r_0^2\approx1$, and with the quench point positioned near the wall at $z_q \approx z_0$. A key finding is that, in contrast to the Poincare AdS case, even massless fields exhibit a distribution in strong agreement with the GSE ensemble (pink line). Let us highlight the important features:

\begin{itemize}
\item The observation of GSE statistics for massless fields is a significant and main departure from the confining AdS case, where chaos was exclusive to heavy fields. This aligns with known results for the spectral form factor (SFF) of BTZ black holes, where normal modes exhibit chaotic behavior for both massive and massless fields~\cite{Das:2023ulz, Ageev:2024gem}.
\item The high-temperature regime of the BTZ black hole strongly enhances chaotic dynamics. We find that at elevated temperatures, the peak ratio distribution lies significantly closer to the GSE prediction than at lower temperatures, indicating a more pronounced level of quantum chaos.
\item In this thermal background, a quite significant number of modes must be taken into account to approach the RMT statistics. This contrasts with the IR-deformed Poincare AdS case, where for certain parameters, as few as $N=5$ series terms were sufficient.
\end{itemize}

\begin{figure}[t!]
\centering
\subfloat[$\nu = 1$, $r_h=1$]{\includegraphics[width=0.5\textwidth]{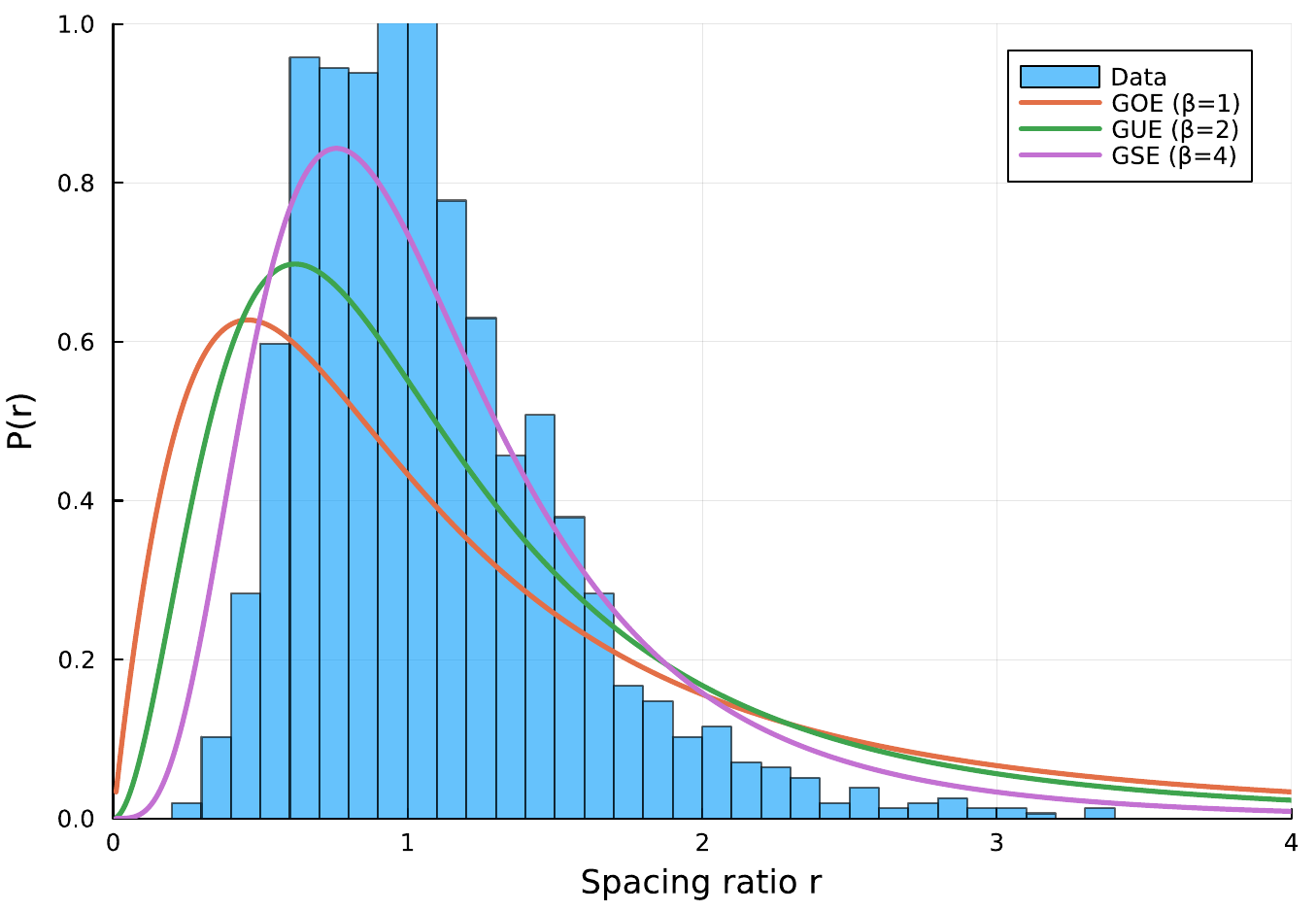}}
\subfloat[$\nu = 3$, $r_h=1$]{\includegraphics[width=0.5\textwidth]{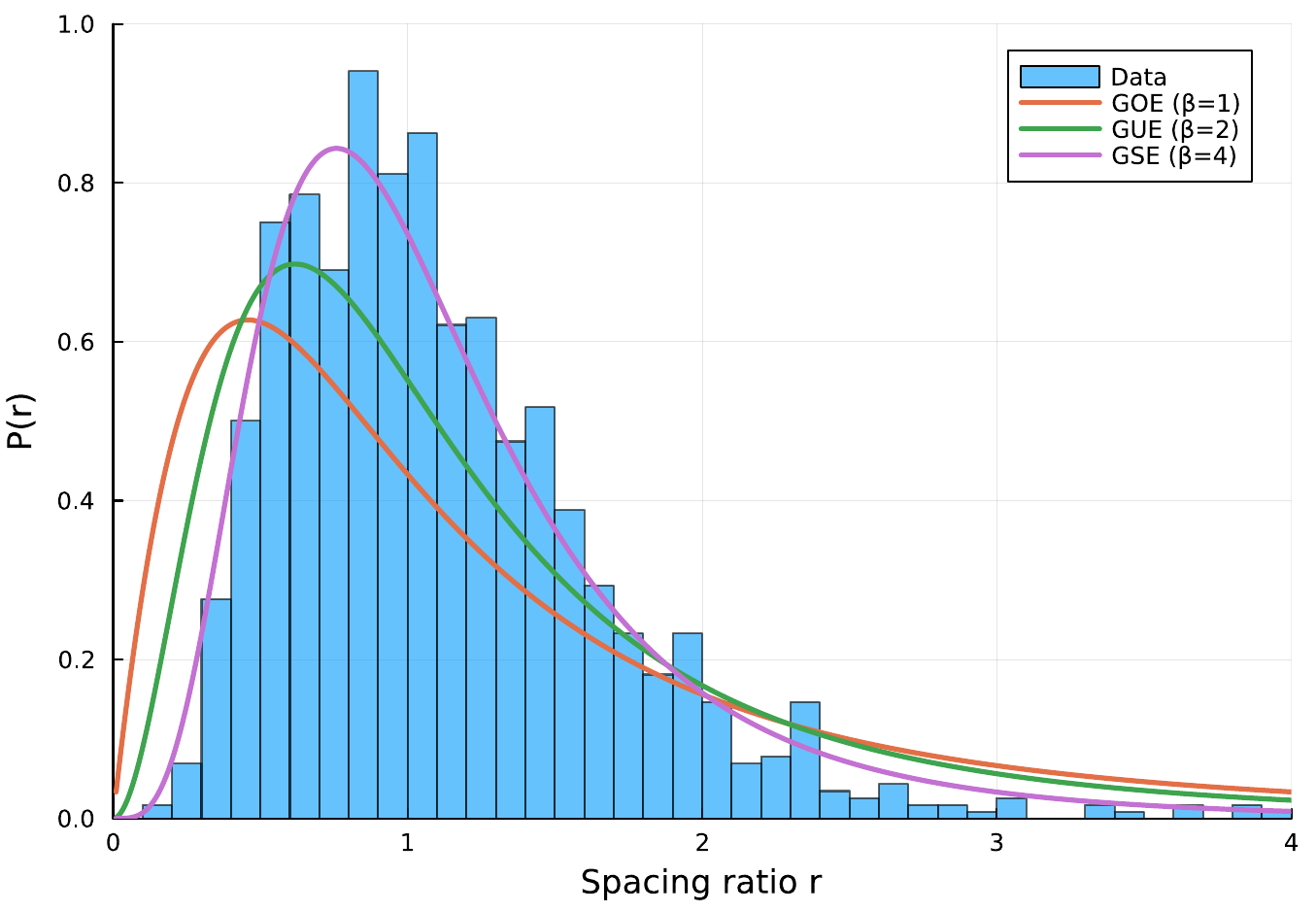}}\\
\subfloat[$\nu = 1$, $r_h=6$]{\includegraphics[width=0.5\textwidth]{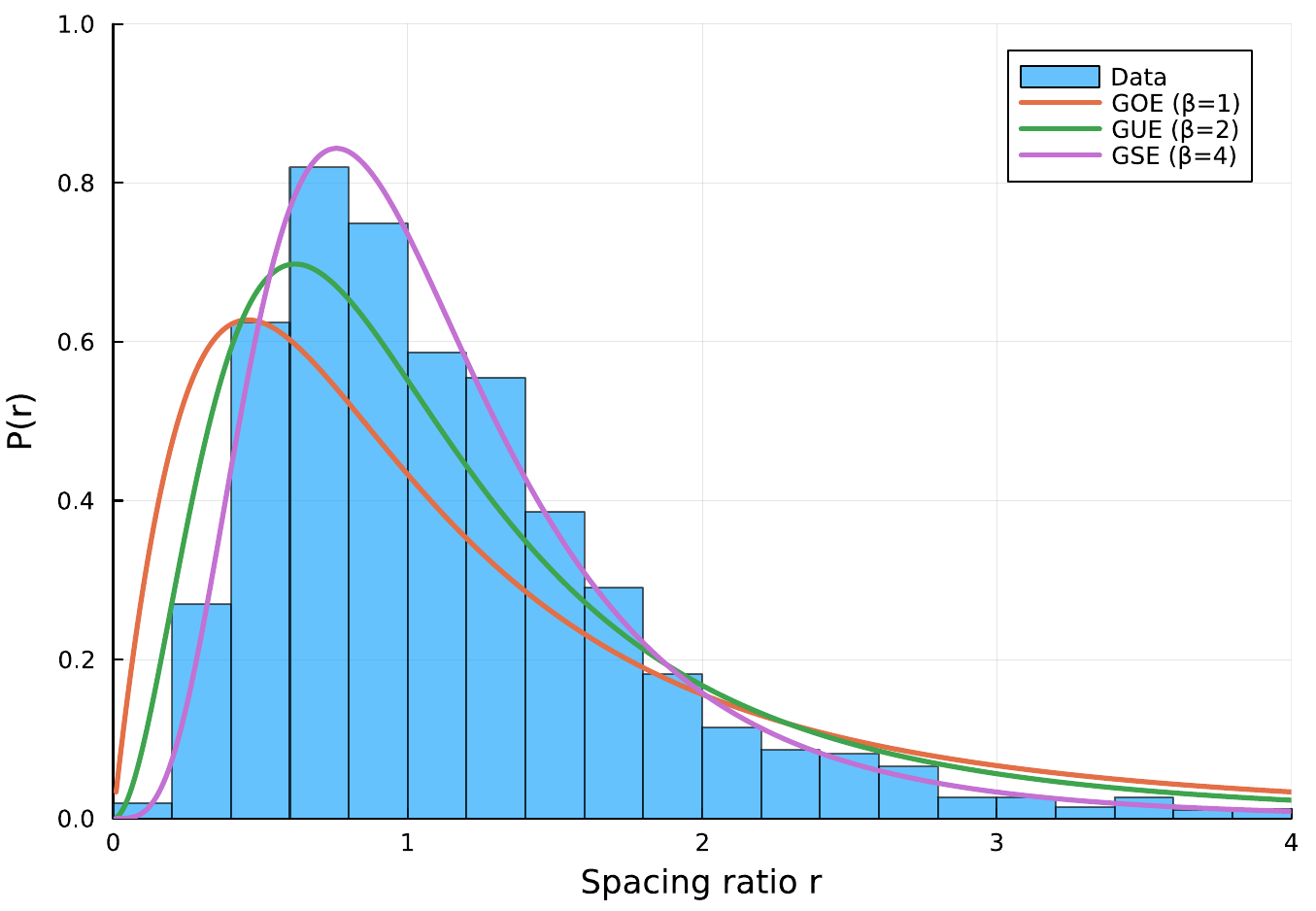}}
\subfloat[$\nu = 3$, $r_h=6$]{\includegraphics[width=0.5\textwidth]{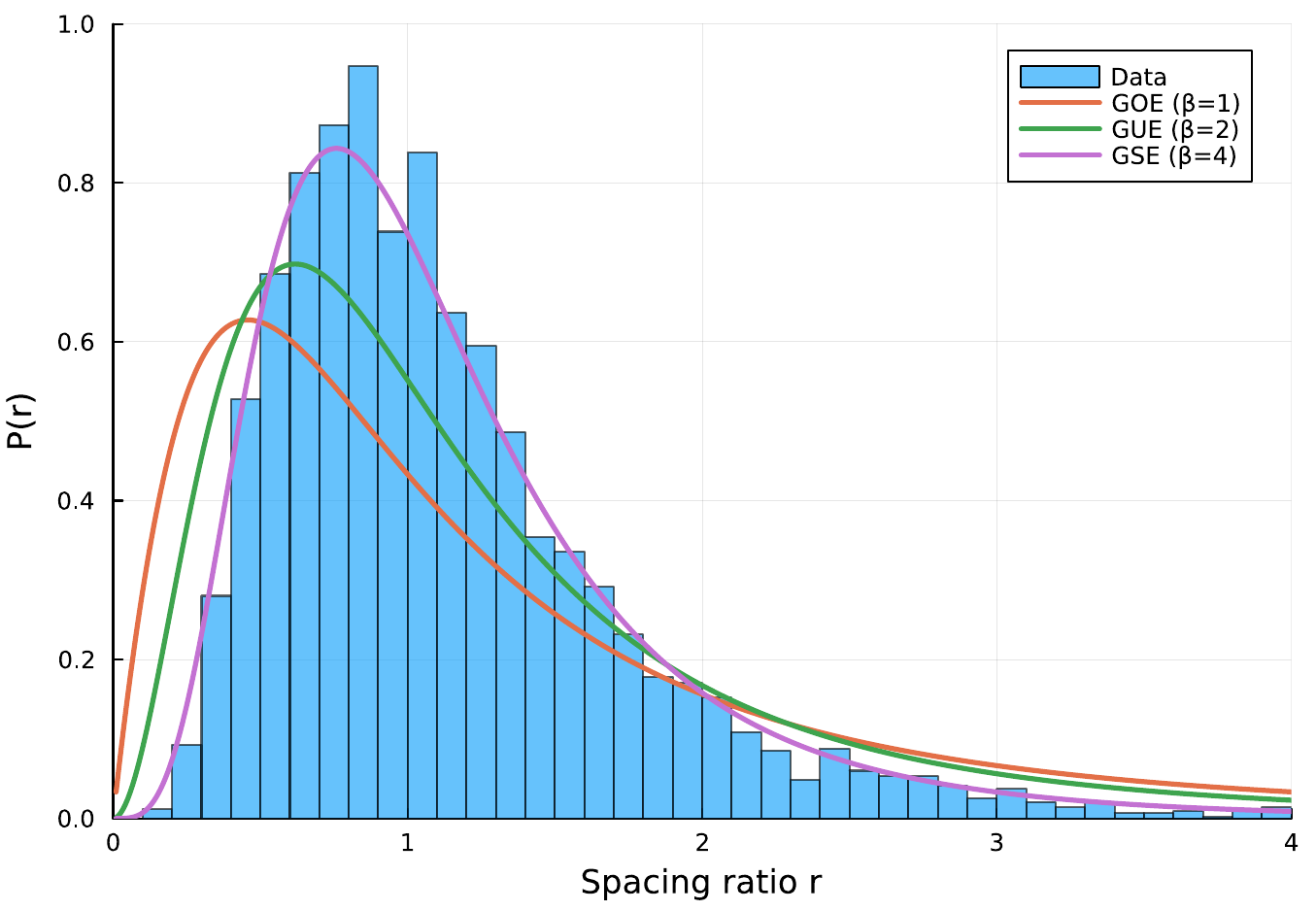}}
\caption{Peak spacing ratio distributions for the boundary operator $\mathcal{O}^2$ in deformed BTZ geometry, demonstrating the persistence of GSE statistics under a distant local quench. The quench point is positioned farther from the wall ($z_q=0.79$) with the same wall placement ($z_0=0.99$). The results shown for both massless ($\nu=1$) and massive ($\nu=3$) fields. Panels (a)-(b) correspond to $r_h=1$, while (c)-(d) show the high-temperature regime ($r_h=6$). Other parameters are fixed as: $\varphi_q=t_q=0$, $\epsilon=0.1$, $L=1$, $N=25$, $J_{\text{max}}=40$.} 
\label{fig:BTZ distribution far quench}
\end{figure}

Following the analysis in deformed Poincare AdS, we initially placed the quench point very close to the brick wall. We now move the quench point farther from the wall, decreasing $z_q$, with the results shown in Fig. \ref{fig:BTZ distribution far quench}. Crucially, we find that chaotic behavior persists. This marks another major difference from the Poincare AdS case, where moving the quench point away from the wall led to a significant decrease in chaos. This resilience underscores that the black hole horizon itself, rather than just the proximity to the wall, is the crucial element driving the chaotic behavior of the system.

\section{Conclusion}
In this paper, we studied the evolution of boundary correlation functions for states excited via a path-integral on AdS spaces (Poincare AdS and compact BTZ) with operator insertions, corresponding to local operator quenches in the bulk.

We deformed the Poincare AdS geometry with an infrared hard wall cutoff, a well-known setup in holographic QCD that induces confinement. In this scenario, we confirmed and extended the findings of \cite{Ageev:2025yiq}, observing a clear trend where the temporal dynamics of boundary one-point observables approach the Gaussian Symplectic Ensemble (GSE) distribution. As in that work, this chaotic behavior occurs exclusively for heavy fields, as indicated by their peak-level spacing statistics, and is sensitive to the quench location relative to the wall.

Our study of the compact BTZ black hole, deformed by a brick wall placed near the horizon, reveals significant departures and establishes a more robust form of quantum chaos. In this thermal background, we demonstrate that even massless fields exhibit statistics consistent with the GSE distribution, a result not observed in the confining case. Crucially, this chaotic behavior persists even when the quench point is moved away from the brick wall, indicating that the horizon itself, rather than just the wall proximity, is the fundamental driver of chaos. Furthermore, the chaotic behavior is markedly temperature dependent. The higher temperature regime drives the level statistics substantially closer to the GSE distribution than at lower temperatures, underscoring the role of thermal excitations in strengthening quantum chaos.

We leave for future work detailed studies of planar and extremal BTZ black holes in this context. Furthermore, it would be interesting to apply recent proposals, such as the generalization of peak statistics to higher-dimensional patterns considered in \cite{Bianchi:2025kna}, to our setup.

\acknowledgments
The work of D. S. Ageev and V.A. Bykov was supported by the Foundation for the Advancement of Theoretical Physics and Mathematics "BASIS" grant \#24-2-1-79-1 (VB) and \#24-1-3-35-1 (DA). The work of D.S. Ageev was performed at the Steklov International Mathematical Center and supported by the Ministry of Science and Higher Education of the Russian Federation (agreement no. 075-15-2025-303).

\appendix
\section{Derivation of Green's function for deformed AdS geometry} \label{App A}
In this section we will derive Green's function for deformed AdS geometry. Green's function $G(y,z; y',z')$ for this case satisfies the following equation
\begin{multline}
    \left(\frac{z^2}{L^2}\partial^2_z +z\frac{1-d}{L^2}\partial_z +\frac{z^2}{L^2}\eta^{\mu\nu}\partial_{\mu}\partial_{\nu}-m^2\right)G(y,z;y',z')=\\
    =-\frac{z^{d+1}}{L^{d+1}}\delta(y-y')\delta(z-z'), \label{green eq ads poincare app}
\end{multline}
with boundary conditions
\begin{gather}
    G(y,z_0;y',z')=G(y,z;y',z_0)=0, \label{green boundary cond ads poincare app} \\
    G(y,0;y',z')=G(y,z;y',0)=0, \label{green boundary cond ads poincare at infty app}
\end{gather}
where $y=(\tau, x)$ is a set of boundary coordinates. The ansatz for $G(y,z;y',z')$ has the form
\be
\label{ansatz for green ads poincare App}
G(y,z;y',z')=\sum_n\int\frac{d^dk}{(2\pi)^d} e^{ik(y-y')}\frac{z^{\frac{d}{2}}z'^{\frac{d}{2}}J_{\nu}(\alpha_n z)J_{\nu}(\alpha_n z')}{N_n}G_n(k),
\ee
where $\alpha_n$ are defined as follows
\be
\label{alpha def wall ads poincare App}
    J_{\nu}(\alpha_n z_0)=0
\ee
to satisfy boundary conditions (\ref{green boundary cond ads poincare app}), the sum goes over all solutions of equation (\ref{alpha def wall ads poincare App})
Our goal is to find such $G_n(k)$ and $N_n$ that (\ref{ansatz for green ads poincare App}) satisfies equation (\ref{green eq ads poincare app}) Substituting the ansatz into equation we get
\begin{multline}
    \sum_n\int\frac{d^dk}{(2\pi)^d}\frac{z'^{\frac{d}{2}}J_{\nu}(\alpha_n z')}{N_n}G_n(k)e^{ik(y-y')}\frac{z^{\frac{d}{2}}}{L^2}(z^2 J_{\nu}''(\alpha_n z)+z J_{\nu}'(\alpha_n z)+\\
    +(-z^2 k^2-m^2 L^2-\frac{d^2}{4})J_{\nu}(\alpha_n z))=-\frac{z^{d+1}}{L^{d+1}}\delta(y-y')\delta(z-z'). \label{J eq}
\end{multline}
Using equation for $J_{\nu}(\alpha_nz)$ we rewrite (\ref{J eq}) as
\begin{multline}
\sum_n\int\frac{d^dk}{(2\pi)^d}\frac{z'^{\frac{d}{2}}J_{\nu}(\alpha_n z')}{N_n}G_n(k)e^{ik(y-y')}\frac{z^{\frac{d}{2}}}{L^2}(-z^2k^2-z^2\alpha_n^2)J_{\nu}(\alpha_n z))=\\
=-\frac{z^{d+1}}{L^{d+1}}\delta(y-y')\delta(z-z').
\end{multline}
From here we assume $G_n(k)$ to be
\be
G_n(k)=\frac{1}{L^{d-1}(k^2+\alpha_n^2)}, \label{Gn ads poincare app}
\ee
so 
\be
z'^{\frac{d}{2}}\int\frac{d^dk}{(2\pi)^d}e^{ik(y-y')}\sum_n\frac{zJ_{\nu}(\alpha_n z)J_{\nu}(\alpha_n z')}{N_n}=z^{\frac{d}{2}}\delta(y-y')\delta(z-z'),
\ee
which is consistent with definition of $\delta(y-y')$ and completeness relation for Bessel functions 
\be
\sum_n\frac{zJ_{\nu}(\alpha_n z))J_{\nu}(\alpha_n z')}{N_n}=\delta(z-z'),
\ee
where 
\be
N_n=\int_0^{z_0}dz zJ_{\nu}^2(\alpha_n z)=\frac{z_0^2}{2}J_{\nu+1}^2(\alpha_n z_0).
\ee
Then Green's function has the form
\be
G(y,z;y',z')=\sum_n\int\frac{d^dk}{(2\pi)^d L^{d-1}} \frac{e^{ik(y-y')}}{k^2+\alpha_n^2}\frac{z^{\frac{d}{2}}z'^{\frac{d}{2}}J_{\nu}(\alpha_n z)J_{\nu}(\alpha_n z')}{N_n}.
\ee
Now let us perform the integration over $k$ in spherical coordinates
\be
\int\frac{d^dk}{(2\pi)^d} \frac{e^{ik(y-y')}}{k^2+\alpha_n^2}=\int_0^{\infty}\frac{k^{d-1}dk}{(2\pi)^d(k^2+\alpha_n^2)}\int_0^{\pi}d\theta e^{ikr\cos{\theta}}\sin^{d-2}{\theta}\Omega_{d-2},
\ee
where $r=\sqrt{(\tau-\tau')^2+(x-x')^2}$ and $\Omega_{d-2}$ is the area of $(d-2)$-dimensional sphere
\be
\Omega_{d-2}=\frac{2\pi^{\frac{d-1}{2}}}{\Gamma(\frac{d-1}{2})}.
\ee
Integration over angle $\theta$ gives us Bessel function 
\be
\int_0^{\pi}d\theta e^{ikr\cos{\theta}}\sin^{d-2}{\theta}=\frac{2^{\frac{d}{2}-1}\sqrt{\pi}\Gamma(\frac{d-1}{2})}{(kr)^{\frac{d}{2}-1}}J_{\frac{d}{2}-1}(kr),
\ee
so we left with
\be
\int\frac{d^dk}{(2\pi)^d} \frac{e^{ik(y-y')}}{k^2+\alpha_n^2}=\frac{1}{(2\pi)^{\frac{d}{2}}r^{\frac{d}{2}-1}}\int_0^{\infty}dk\frac{k^{\frac{d}{2}}}{k^2+\alpha_n^2}J_{\frac{d}{2}-1}(kr).
\ee
The last integral is equal to
\be
\int_0^{\infty}dk\frac{k^{\frac{d}{2}}}{k^2+\alpha_n^2}J_{\frac{d}{2}-1}(kr)=\alpha_n^{\frac{d}{2}-1}K_{\frac{d}{2}-1}(\alpha_n r).
\ee
The final result for Green's function is given by
\be
G(\tau,x,z;\tau',x',z')=\sum_n\left(\frac{\alpha_n}{r}\right)^{\frac{d}{2}-1}\frac{z^{\frac{d}{2}}z'^{\frac{d}{2}}J_{\nu}(\alpha_n z)J_{\nu}(\alpha_n z')}{(2\pi)^{\frac{d}{2}}L^{d-1}N_n}K_{\frac{d}{2}-1}(\alpha_n r).
\ee
\section{Derivation of Green's function for deformed BTZ geometry} \label{App B}
In this section we will derive Green's function for deformed BTZ geometry. Green's function $G_{DBTZ}(\tau,\varphi,r;\tau',\varphi',r')$ for this case satisfies the following equation
\begin{multline}
    \left(\frac{r^2-r_h^2}{L^2}\partial^2_r +\frac{3r^2-r_h^2}{rL^2}\partial_r +\frac{L^2}{r^2-r_h^2}\partial^2_{\tau}+\frac{1}{r^2}\partial^2_{\varphi}-m^2\right)G_{DBTZ}=\\
    =-\frac{1}{r}\delta(\tau-\tau')\delta(\varphi-\varphi')\delta(r-r'), \label{green eq BTZ app}
\end{multline}
with boundary conditions
\begin{gather}
    G_{DBTZ}(\tau,\varphi,r_0;\tau',\varphi',r')=G_{DBTZ}(\tau,\varphi,r;\tau',\varphi',r_0)=0 \label{boundary cond for green BTZ app},\\
     G_{DBTZ}(\tau,\varphi,\infty;\tau',\varphi',r')=G_{DBTZ}(\tau,\varphi,r;\tau',\varphi',\infty)=0 \label{boundary cond for green BTZ app at infty},\\
    G_{DBTZ}(\tau,\varphi+2\pi n,r;\tau',\varphi',r')=G_{DBTZ}(\tau,\varphi,r;\tau',\varphi',r') \label{periodicity cond app}
\end{gather}
Ansatz for $G_{DBTZ}(\tau,\varphi,r;\tau',\varphi',r')$ has the form
\be
\label{ansatz for green BTZ App}
G_{DBTZ}(\tau,\varphi,r;\tau',\varphi',r')=\sum_n\sum_J\int d\lambda e^{i\lambda(\tau-\tau')}e^{iJ(\varphi-\varphi')}\frac{F_{n,J}(r)F_{n,J}(r')}{(2\pi)^2N_{n,J}\sqrt{rr'}}G_{n,J}(\lambda),
\ee
where $J$ is integer and index $n$ numerates a set of solutions to equation
\be
\label{boundary cond F}
F_{n,J}(r_0)=0,
\ee
enforcing the boundary condition (\ref{boundary cond for green BTZ app}). More comments on that later. Substituting the ansatz into equation we get
\begin{multline}
   \sum_n\sum_J\int d\lambda e^{i\lambda(\tau-\tau')}e^{iJ(\varphi-\varphi')}\frac{F_{n,J}(r')}{(2\pi)^2 N_{n,J}\sqrt{rr'}L^2}G_{n,J}(\lambda)\bigg[(r^2-r_h^2)\partial^2_r+2r\partial_r
   -\bigg(\frac{r_h^2}{4r^2}+\\
   +\frac{3}{4}+\frac{L^2J^2}{r^2}+m^2L^2\bigg)
   -\frac{L^4\lambda^2}{r^2-r^2_h}\bigg]F_{n,J}(r)=-\frac{1}{r}\delta(\tau-\tau')\delta(\varphi-\varphi')\delta(r-r'). \label{Ansat eq BTZ}
\end{multline}
We pick function $F_{n,J}(r)$ as a solution to equation
\be
\left[(r^2-r_h^2)\partial^2_r+2r\partial_r-\left(\frac{r_h^2}{4r^2}+\frac{3}{4}+\frac{L^2J^2}{r^2}+m^2L^2\right)+\frac{L^4\omega_{n,J}^2}{r^2-r^2_h}\right]F_{n,J}(r)=0. \label{F eq}
\ee
Then equation (\ref{Ansat eq BTZ}) takes form
\begin{multline}
     \sum_n\sum_J\int\frac{d\lambda}{(2\pi)^2} e^{i\lambda(\tau-\tau')}e^{iJ(\varphi-\varphi')}\frac{1}{\sqrt{rr'}}\frac{F_{n,J}(r')}{N_{n,J}}G_{n,J}(\lambda)\frac{L^2(\lambda^2+\omega_{n,J}^2)}{r^2-r^2_h}F_{n,J}(r)=\\
     =\frac{1}{r}\delta(\tau-\tau')\delta(\varphi-\varphi')\delta(r-r') \label{eq and deltas}
\end{multline}
From here we assume $G_{n,J}(\lambda)$ to be
\be
G_{n,J}(\lambda)=\frac{2L^2}{r_h(\lambda^2+\omega_{n,J}^2)}, \label{GnJ BTZ app}
\ee
so (\ref{eq and deltas}) takes form
\begin{multline}
 \frac{1}{\sqrt{rr'}}\sum_n\sum_J\int\frac{d\lambda}{(2\pi)^2} e^{i\lambda(\tau-\tau')}e^{iJ(\varphi-\varphi')}\frac{F_{n,J}(r)F_{n,J}(r')}{N_{n,J}}\frac{2L^4}{(r^2-r^2_h)r_h}=\\
 =\frac{1}{r}\delta(\tau-\tau')\delta(\varphi-\varphi')\delta(r-r'). \label{delta identities}
\end{multline}
Next we analyze equation (\ref{F eq}). Taking into account the boundary condition (\ref{boundary cond for green BTZ app at infty}), the solution is given by
\be
 F_{n,J}(r)=\left(1-\frac{r_h^2}{r^2}\right)^{\alpha_n}\left(\frac{r_h^2}{r^2}\right)^{\beta}{_2}F_1\left(a_{n,J},b_{n,J};c;\frac{r_h^2}{r^2}\right),
\ee
where parameters $a_n$, $b_n$, $c$, $\alpha_n$ and $\beta$ are equal to
\begin{gather}
    a_n,b_n=\frac{1}{2}\left(1+\nu+\frac{iL^2\omega_n}{r_h}\mp\frac{iLJ}{r_h}\right),  \quad c=\Delta \label{hypergeometric param app},\\
    \alpha_n=\frac{iL^2\omega_n}{2r_h}, \quad \beta=\frac{1}{4}+\frac{\nu}{2}. \label{alpha and beta app}
\end{gather}
To obtain completeness relation for functions $F_{n,J}(r)$ let us introduce coordinates $z=r_h^2/r^2$ and ansatz $F_{n,J}(z)=z^{3/4}f_{n,J}(z)$. Then equation (\ref{F eq}) takes form
\begin{multline}
4z^2(1-z)f''_{n,J}(z)+4z(2-3z)f'_{n,J}(z)-\left(4z+m^2L^2+\frac{L^2J^2z}{r_h^2}\right)f_{n,J}(z)+\\
+\frac{L^4z}{r_h^2(1-z)}\omega^2_{n,J}f_{n,J}(z)=0, \label{f eq}
\end{multline}
with boundary conditions 
\be
\label{boundary conditions f}
f_{n,J}(z_0)=f_{n,J}(0)=0,
\ee
which are just rewritten conditions (\ref{boundary cond for green BTZ app at infty}) and (\ref{boundary cond F}). Equations (\ref{f eq}-\ref{boundary conditions f}) define a regular Sturm-Liouville problem with eigenvalue $\omega^2_{n,J}$ and weight factor 
\be
\label{weight}
\rho(z)=\frac{L^4z}{r_h^2(1-z)},
\ee
where eigenvalues $\omega_{n,J}$ are defined as the solutions to boundary equation (\ref{boundary cond F}), which may be expressed as
\be
{_2}F_1(a_{n,J},b_{n,J};c;z_0)=0.
\ee
Functions $f_{n,J}(z)$, as solutions to a Sturm-Liouville problem, form a complete set of functions, with completeness relation
\be
\sum_n\rho(z)\frac{f_{n,J}(z)f_{n,J}(z')}{N_{n,J}}=\delta(z-z'), \label{delta z}
\ee
where normalisation factors $N_{n,J}$
are defined as
\be
N_{n,J}=\int_0^{z_0}\rho(z)f^2_{n,J}(z)dz=\frac{L^4}{r_h^2}\int_{0}^{z_0}(1-z)^{2\alpha_{n,J}-1}z^{\nu}{_2}F_1^2(a_{n,J},b_{n,J};c;z)dz
\ee
Let us return to r coordinates, using 
\be
\delta(z-z')=\frac{\delta(r-r')}{|\frac{dz}{dr}|}=\frac{\delta(r-r')}{\frac{2r_h^2}{r^3}}.
\ee
Than (\ref{delta z}) takes form
\be
\delta(r-r')=\sum_n\frac{F_{n,J}(r)F_{n,J}(r')}{N_{n,J}}\frac{2L^4}{(r^2-r^2_h)r_h}, \label{delta r}
\ee
so expression (\ref{delta identities}) is consistent with (\ref{delta r}) and basic definitions of $\delta(\tau-\tau')$ and $\delta(\varphi-\varphi')$, which concludes our derivation.

\end{document}